\documentclass[3p]{elsarticle}

\usepackage{url}
\usepackage{bm}
\usepackage{amssymb}
\usepackage{amsmath}
\usepackage{amsthm}
\usepackage{color}
\usepackage{subfigure}
\usepackage{booktabs}
\usepackage{multirow}

\biboptions{sort&compress}
\newdefinition{remark}{Remark}

\def\abs #1{\left|#1\right|}

\def\st{\text{subject to }}

\def\bC{\mathbb{C}}

\def\bS{\mathbb{S}}

\def\m #1{\boldsymbol{#1}}

\def\cC{\mathcal{C}}

\def\cH{\mathcal{H}}

\def\bee{\begin{equation}}
	\def\ene{\end{equation}}

\def\beq{\begin{eqnarray}}
	\def\enq{\end{eqnarray}}
\def\lentwo{\setlength\arraycolsep{2pt}}

\newtheorem{lem}{Lemma}

\newtheorem{thm}{Theorem}

\def\equ #1{\begin{equation}#1\end{equation}}
\def\equa #1{\begin{eqnarray}#1\end{eqnarray}}
\def\sbra #1{\left(#1\right)}
\def\mbra #1{\left[#1\right]}
\def\lbra #1{\left\{#1\right\}}

\def\tr #1{\text{tr}#1}

\def\rank #1{\text{rank}#1}
\def\st {\text{ subject to }}

\DeclareMathOperator*{\argmin}{arg\,min}

\begin{document}
	
\begin{frontmatter}

\title{Pursuing the Limit of Chirp Parameter Identifiability:\\ A Computational Approach}

\author[1]{Zai~Yang}
\ead{yangzai@xjtu.edu.cn}
\author[1]{Sikai~Ge}
\ead{gesikai010421@stu.xjtu.edu.cn}
\author[1]{Wenlong~Wang}
\ead{wang1813857265@stu.xjtu.edu.cn}
\affiliation[1]{organization={School of Mathematics and Statistics, Xi'an Jiaotong University},
city={Xi'an, Shaanxi 710049},
country={China}}

\begin{abstract}
In this paper, it is shown that a necessary condition for unique identifiability of $K$ chirps from $N$ regularly spaced samples of their mixture is $N\geq 2K$ when $K\geq 2$. A necessary and sufficient condition is that a rank-constrained matrix optimization problem has a unique solution; this is the first result of such kind. An algorithm is proposed to solve the optimization problem and to identify the parameters numerically. The lower bound of $N=2K$ is shown to be tight by providing diverse problem instances for which the proposed algorithm succeeds to identify the parameters. The advantageous performance of the proposed algorithm is also demonstrated compared with the state of the art.
\end{abstract}



\end{frontmatter}

\section{Introduction}
A (linear) chirp is a signal whose frequency changes linearly with time. It has been widely applied in radar, sonar and communication systems due to its great pulse compression capacity \cite{klauder1960theory,springer2000spread,burgess1992chirp,kronauge2014new}. Motivated by these applications, a fundamental problem in signal processing and harmonic analysis is to separate a number of chirps from discrete-time samples of their mixtures. Since a continuous-time chirp is uniquely determined by its complex amplitude, initial frequency and chirp rate, the above problem is also known as chirp parameter estimation. 

The topic of chirp parameter estimation has been extensively studied. One kind of prevalent approaches is based on time--frequency representation, which is initially developed for blind source separation and does not rely on an explicit data model. The short-time Fourier transform (STFT) \cite{griffin1984signal,durak2003short,tao2009short} enables to estimate the instantaneous frequency of the signal by applying the Fourier transform (FT) to the signal within a short time window. Due to the resolution limit of FT, STFT needs to balance between the resolution in frequency and time and suffers from poor performance when the chirp components share cross instantaneous frequencies. The Wigner-Ville distribution (WVD) \cite{debnath2002wigner} enhances the time--frequency resolution by applying the FT to the signal's symmetric instantaneous autocorrelation function (SIAF), but the non-coherent accumulation of cross terms among chirp components degrades its performance. As an extension of STFT, chirplet transform \cite{mann1995chirplet} can analyze how rapidly the frequency changes by adding a second-order time-related variable. It is shown in \cite{bennett2005using} that using edge information extracted from time--frequency representations can enhance the performance of chirp parameter estimation. Since the explicit chirp data model is not utilized in these methods, sufficient samples are required to separate the chirps. Moreover, a Nyquist sampling rate is required in these methods to avoid aliasing of frequencies.

Another kind of methods partially utilizes the chirp data model. The continuous fractional Fourier transform (CFRFT) \cite{bailey1991fractional,sejdic2011fractional,aldimashki2020performance} generalizes the conventional FT and introduces an arbitrary rotation angle associated with the chirp rate. The chirp parameters can be estimated from the peaks of the spectrum in the two-dimensional (2-D) FRFT domain. The discrete FRFT (DFRFT) \cite{ozaktas1996digital,candan2000discrete} is an approximate realization of the CFRFT given discrete-time observations. The discrete chirp Fourier transform (DCFT) \cite{xia2000discrete,fan2000modified} shares conceptual similarities with the discrete Fourier transform (DFT). It introduces another degree of freedom to the DFT, which aligns with the chirp rate, and achieves 2-D coherent accumulation in the chirp rate--initial frequency domain. The Lv's transform (LVT) \cite{lv2011lv}, which is composed of a scaling operation and a 2-D DFT, achieves improved coherent accumulation in the 2-D centroid frequency--chirp rate domain and enhances the mainlobe-to-sidelobe ratio as compared to the DCFT. In general, this kind of methods can be viewed as 2-D extensions of FT tailored for chirps and thus suffer from low resolution with limited samples due to  spectral leakage of FT. 

Inspired by the compressed sensing theory \cite{candes2006robust,donoho2006compressed}, methods based on sparse representation have also been proposed for chirp parameter estimation \cite{guo2011parameter,sward2015sparse,al2019compressive,yang2025gridless}. This type of methods attempts to select a small number of chirp components from an overcomplete dictionary by solving a sparse optimization problem. For instance, the authors in \cite{al2019compressive} use DCFT basis to generate a sparse representation of the multicomponent chirp signal, and then an $\ell_1$ norm minimization problem is proposed and approximately solved by orthogonal matching pursuit (OMP) \cite{pati1993orthogonal}. However, these optimization methods usually rely on parameter discretization and problem relaxation and thus cannot exploit the full structures of the data model, resulting in suboptimal solutions.

While previous studies have mainly been focused on improving algorithm performance for chirp parameter estimation, a fundamental question has been unresolved as to how many regularly spaced samples are necessary and sufficient for chirp parameter identifiability in the ideal noiseless setting. This paper is devoted to answering this question from a computational perspective. Our main findings are summarized below.
\begin{itemize}
\item Three samples are necessary and sufficient to uniquely identify a single chirp under mild conditions.
\item $N\geq 2K$ samples are necessary to identify $K\geq 2$ chirps, and the lower bound $N=2K$ is attainable and thus tight.
\item In order that the chirp parameters are uniquely identifiable, a necessary and sufficient condition is that a rank-constrained matrix feasibility problem has a unique solution; this is the first result of such kind. 
\item An algorithm is proposed to solve the matrix optimization problem by semidefinite programming, which enables exact parameter recovery from the minimal $N=2K$ samples across diverse problem instances, thereby empirically verifying the tightness of the lower bound and the superior performance of the proposed algorithm  over existing methods.
\end{itemize}

The rest of the paper is organized as follows. The problem description of chirp parameter identifiability is introduced in Section \ref{sec:problem}. Necessary and sufficient conditions for unique identifiability are presented in Section \ref{sec:theory}. An algorithm for parameter recovery is proposed in Section \ref{sec:algorithm}. Numerical results are provided in Section \ref{sec:numerical}. Conclusions are finally drawn in Section \ref{sec:conclusion}.

\section{Problem Description} \label{sec:problem}
Consider a continuous-time multicomponent signal $y(t)$ composed of $K$ chirp signals and given by: 
\equ{y(t)=\sum_{k=1}^K s_k e^{i2\pi(f_{k}^o t+\tau_{k}^o t^2)},  \ \ 0 \leq t \leq T, \label{1}}
where $K$ is the number of chirps, $s_k$ is the complex amplitude of the $k$th chirp signal, $f_{k}^o$ denotes the initial frequency of the $k$th chirp, $\tau_{k}^o$ is the chirp rate, and $T$ is the time duration. Note that the instantaneous frequency of the $k$th chirp at time $t$ is $f_k^o+2\tau_k^o t$, the derivative of the phase function $f_{k}^o t+\tau_{k}^o t^2$, and the true chirp rate is $2\tau_k^o$, where the scaling factor of 2 is omitted in this paper for convenience. Sampling $\m{y}(t)$ regularly at a rate $f_s$ gives the equispaced discrete-time samples
\equ{{y}[n]=\sum_{k=1}^K s_k e^{i2\pi(f_{k}n+\tau_kn^2)},  \ \ 0 \leq n \leq N-1, \label{2}}
where $f_k=\frac{f_{k}^o}{f_s}$, $\tau_k=\frac{\tau_{k}^o}{f_s^2}$ denote the {\em normalized} initial frequency and chirp rate, respectively, and $N = {T}{f_s}+1$ is the sample size. Our objective is to identify/recover the parameters $\lbra{s_k, f_k, \tau_k}_{k=1}^K$ given the length-$N$ vector $\m{y}$ of samples.

If the samples are acquired at a Nyquist sampling rate, then we have the initial frequencies $\abs{f_k}\leq \frac{1}{2}$ and the final frequencies $\abs{f_k + 2\tau_k (N-1)} \leq \frac{1}{2}$, yielding that $\abs{2\tau_k (N-1)} \leq 1$ and thus $\abs{\tau_k} \leq \frac{1}{2(N-1)}$. In this paper, we will assume that $\abs{f_k}\leq \frac{1}{2}$ and $\abs{\tau_k}\leq \delta<\frac{1}{2}$, where $\delta>0$ is given {\em a priori}. Although the assumption $\abs{\tau_k}\leq \delta$ plays an important role for parameter identifiability, we will show that a Nyquist sampling rate is unnecessary.

\section{Theory} \label{sec:theory}
We first show the following result, which follows from the principle that the number of unknowns cannot exceed the number of measurements.
\begin{thm}
	If the parameters $\lbra{s_k, f_k, \tau_k}_{k=1}^K$ can be uniquely identified from the length-$N$ vector $\m{y}$ given in \eqref{2}, then $N\geq 2K$. \label{thm:N2K}
\end{thm}

\begin{proof}
	We apply the implicit function theorem to prove the theorem. To do so, we stack the parameters $\lbra{s_k}_{k=1}^K$, $\lbra{f_k}_{k=1}^K$, and $\lbra{\tau_k}_{k=1}^K$ into the following vectors:
\begin{equation}
\m{f}=\left[f_1,\cdots,f_K\right]^\top \in \mathbb{R}^{K},\quad \m{\tau}=\left[\tau_1,\cdots,\tau_K\right]^\top \in \mathbb{R}^{K},\quad \m{s}=\left[s_1,\cdots,s_K\right]^\top \in \mathbb{C}^{K}. \notag
\end{equation}
	We define a real-valued parameter vector as
\begin{equation}
\m{v}=\left[\m{f}^\top,\m{\tau}^\top,\Re\left\{\m{s}\right\}^\top,\Im\left\{\m{s}\right\}^\top\right]^\top\in\mathbb{R}^{4K}.\notag
\end{equation}
Similarly, we stack the real and imaginary parts of the observations into the real vector:
	\begin{equation}
		\begin{aligned}
			\widetilde{\m{y}}=\left[\Re\left\{y\left[1\right]\right\},\cdots,\Re\left\{y\left[N\right]\right\},\Im\left\{y\left[1\right]\right\},\cdots,\Im\left\{y\left[N\right]\right\}\right]^\top\in\mathbb{R}^{2N}.
		\end{aligned}\notag
	\end{equation}
	Define the smooth function representing the observation system as
	\begin{equation}
		\begin{aligned}
			F\left(\m{v}\right)=\widetilde{\m{y}}.
		\end{aligned}\notag
	\end{equation}
	Suppose there exists a point $\m{v}_0$ such that the Jacobian matrix $\frac{\partial}{\partial\m{v}}F$ evaluated at $\m{v}_0$ has rank $r<4K$. Then, by the implicit function theorem, there must exist an open set $G \subseteq \mathbb{R}^{4K-r}$ and a holomorphic map $h$ such that for any $\m{g} \in G$, if $\m{v}=\left[\m{g}^\top,h\left(\m{g}\right)^\top\right]^\top$, then $F\left(\m{v}\right)=F\left(\m{v}_0\right)$, implying that there exist infinitely many solutions that are indistinguishable from $\m{v}_0$.
	
	Note that the Jacobian $\frac{\partial}{\partial\m{v}}F\in\mathbb{R}^{2N\times 4K}$ can have rank at most $2N$. It follows that if $N<2K$, then
	\begin{equation}
		\begin{aligned}
			\rank\left(\frac{\partial}{\partial\m{v}}F\right)\leq2N<4K,
		\end{aligned}\notag
	\end{equation}
	which implies that any point $\m{v}$ cannot be uniquely identified. Therefore, the condition $N\geq 2K$ is necessary for unique identifiability of the parameters $\lbra{s_k, f_k, \tau_k}_{k=1}^K$, completing the proof.
\end{proof}

\begin{thm} Given $K=1$, a necessary condition for unique identifiability of the parameters $\lbra{f_1,\tau_1,s_1}$ is $N\geq 3$. If, further, $\abs{f_1} < \frac{1}{4}$ and $\delta<\frac{1}{4}$, then $N\geq 3$ is also sufficient. \label{thm:K1}
\end{thm}
\begin{proof}
If $\lbra{f_1,\tau_1,s_1}$ are uniquely identifiable, by Theorem \ref{thm:N2K}, we have $N\geq 2$. To conclude $N\geq 3$, it suffices to show that $N=2$ is insufficient. In particular, we have in this case 
\equ{y[0] = s_1,\quad y[1]=s_1e^{i2\pi(f_1+\tau_1)}, \label{eq:y01}}
and thus it is only possible to identify $f_1+\tau_1$ rather than $f_1,\tau_1$ individually.

When $N\geq 3$, we have
\equ{y[0] = s_1,\quad y[1]=s_1e^{i2\pi(f_1+\tau_1)},\quad y[2]=s_1e^{i2\pi(2f_1+4\tau_1)}.\notag}
It follows immediately that
\equ{s_1 = y[0],\quad e^{i4\pi f_1} =(y[0])^{-3} (y[1])^{4} y[2], \quad e^{i4\pi\tau_1} =y[0] (y[1])^{-2} y[2],\notag}
which leads to unique identifiability of $\lbra{f_1,\tau_1,s_1}$ given $\abs{f_1} < \frac{1}{4}$ and $\abs{\tau_1} < \frac{1}{4}$, completing the proof.
\end{proof}

Theorem \ref{thm:K1} shows that the lower bound $N=2K$ of Theorem \ref{thm:N2K} is non-tight as $K=1$. An explanation is that, according to \eqref{eq:y01}, $y[1]$ differs from $y[0]$ only by the phase and these two samples have only 3 (rather than 4) degrees of freedom in total. It is natural to ask whether the lower bound is tight for $K\geq 2$, which will be tested numerically in Section \ref{sec:numerical}.

Since taking more samples at a fixed sampling rate (by increasing the sampling duration) does not weaken the parameter identifiability, we have immediately the following result.

\begin{thm}
	If the parameters $\lbra{s_k, f_k, \tau_k}_{k=1}^K$ can be uniquely identified from $\lbra{y[n]}_{n=0}^{N_1-1}$ given in \eqref{2}, then it can be uniquely identified from $\lbra{y[n]}_{n=0}^{N-1}$ if $N>N_1$. \label{thm:largeN}
\end{thm}

Given some $\tau_k\neq 0$ and sufficiently large $N$, it must hold that $\abs{\tau_k} > \frac{1}{2(N-1)}$ or $\abs{2(N-1)\tau_k} > 1$, meaning that the samples $\lbra{y[n]}_{n=0}^{N-1}$ must be acquired at a sub-Nyquist rate, according to the discussions in Section \ref{sec:problem}. Therefore, it is implied by Theorem \ref{thm:largeN} that a Nyquist sampling rate is unnecessary for chirp parameter identifiability.

Before presenting the next result, we introduce some notations. Given odd numbers $N_1,N_2$ and an $N_1\times N_2$ matrix $\m{Y}$, $\cH\m{Y}$ denotes a $\sbra{\frac{N_1+1}{2}, \frac{N_2+1}{2}}$, 2-level Hankel (or Hankel-block-Hankel) matrix whose $(j,l)$ block, $0\leq j,l\leq \frac{N_1+1}{2}$, is a $\frac{N_2+1}{2} \times \frac{N_2+1}{2}$ Hankel matrix whose $(p,q)$ entry is given by $Y[j+l,p+q]$, where $0\leq p,q\leq \frac{N_2+1}{2}$. Note that $\cH\m{Y}$ is of size $\frac{(N_1+1)(N_2+1)}{4} \times \frac{(N_1+1)(N_2+1)}{4}$. Similarly, a $\sbra{\frac{N_1+1}{2}, \frac{N_2+1}{2}}$, 2-level Toeplitz (or Toeplitz-block-Toeplitz) matrix is $\frac{N_1+1}{2} \times \frac{N_1+1}{2}$ block Toeplitz whose $(j,l)$ block, depending only on $j-l$, is $\frac{N_2+1}{2} \times \frac{N_2+1}{2}$ Toeplitz whose $(p,q)$ entry depends only on $p-q$, where $0\leq j,l\leq \frac{N_1+1}{2}$ and $0\leq p,q\leq \frac{N_2+1}{2}$. Moreover, let $\bS_+^K$ denote the set of positive-semidefinite matrices of rank on greater than $K$. That matrix $\m{A}$ is positive-semidefinite is 
expressed as $\m{A}\geq \m{0}$.

\begin{thm}
	The parameters $\lbra{s_k, f_k, \tau_k}_{k=1}^K$, where $\abs{f_k}\leq \frac{1}{2}$ and $\abs{\tau_k}\leq \delta<\frac{1}{2}$, can be uniquely identified from the length-$N$ vector $\m{y}$ given in \eqref{2} if and only if there exist a unique $M\times \mbra{(M-1)^2+1}$ matrix $\m{Y}$ and a unique $\sbra{\frac{M+1}{2}, \frac{(M-1)^2+2}{2}}$, 2-level Toeplitz matrix $\m{T}$ satisfying that 
	\equa{
		&&\begin{bmatrix} \overline{\m{T}} & \overline{\cH\m{Y}} \\ \cH\m{Y} & \m{T}\end{bmatrix} \in \bS_+^K, \label{eq:HTLMI} \\
		&& \m{T}_{\delta} \geq\m{0}, \label{eq:Tgpsd} \\
		&& Y[n,n^2] = y[n], \quad n=0,\dots,N-1, \label{eq:Yconstr}}
	where $M\geq \max\lbra{N,2K+1}$ is odd, and $\m{T}_{\delta}$ is a $\frac{(M-1)^2}{2}\times \frac{(M-1)^2}{2}$ Toeplitz matrix with
	\equ{T_{\delta}[p,q] = T[p,q+1]+2\cos(2\delta \pi)T[p,q]+T[p+1,q],\quad 0\leq p,q\leq \frac{(M-1)^2-2}{2}.\notag}  \label{thm:equiv}
\end{thm}

To prove Theorem \ref{thm:equiv}, we will use the following results taking from our previous works \cite{wu2022maximum,yang2018frequency,yang2016vandermonde}.

\begin{lem}
	For odd numbers $N_1, N_2$ and $K < \min\lbra{\frac{N_1+1}{2},\frac{N_2+1}{2}}$, the parameters $\lbra{s_k, f_k, \tau_k}_{k=1}^K$, where $\abs{f_k}\leq \frac{1}{2}$ and $\abs{\tau_k}\leq \frac{1}{2}$, can be uniquely identified from the $N_1\times N_2$ matrix $\m{Y}$ given by 
	\equ{Y[j,l] = \sum_{k=1}^K s_k e^{i2\pi (j f_{k}+l\tau_k)}, \quad j=0,\dots,N_1-1,\quad l=0,\dots,N_2-1, \label{eq:Yexpression}} 
	if and only if there exists a unique 2-level Toeplitz matrix $\m{T}$ satisfying \eqref{eq:HTLMI}, where $\m{T}$ is $\frac{N_1+1}{2}\times \frac{N_1+1}{2}$ block Toeplitz with each block being $\frac{N_2+1}{2}\times \frac{N_2+1}{2}$ Toeplitz. Moreover, the $(p,q)$ entry of the $(j,l)$ block of $\m{T}$ must be given by 
	\equ{T_{jl}[p,q] = \sum_{k=1}^K \abs{s_k} e^{i2\pi[(j-l)f_k+(p-q)\tau_k]}. \label{eq:Texpression}} \label{lem:embedding}
\end{lem}
\begin{proof}
See \cite[Corollary 1]{wu2022maximum}.
\end{proof}

\begin{lem}
	Given $-\frac{1}{2}\leq a<b\leq \frac{1}{2}$, an $N\times N$ Hermitian Toeplitz matrix $\m{T}$ admits the expression
	\equ{T[p,q] = \sum_{k=1}^r p_k e^{i2\pi(p-q)f_k},\quad 0\leq p,q\leq N-1,\notag}
	where $p_k>0$, $f_k\in[a,b]$, and $r=\rank\sbra{\m{T}}$, if and only if both $\m{T}$ and $\m{T}_{[a,b]}$ are positive semidefinite, where $\m{T}_{[a,b]}$ is $(N-1)\times (N-1)$ Toeplitz given by
	\equ{T_{[a,b]}[p,q] = e^{i\pi (a+b)} T[p,q+1]-2\cos(\pi(b-a))T[p,q]+ e^{-i\pi (a+b)}T[p+1,q],\quad 0\leq p,q\leq N-1.\notag} 
	Moreover, the set of parameters $\lbra{p_k, f_k}_{k=1}^K$ is unique if either $\m{T}$ or $\m{T}_{[a,b]}$ is rank-deficient.
	\label{lem:selective}
\end{lem}
\begin{proof}
See \cite[Theorem 2]{yang2018frequency}.
\end{proof}

Now we are ready to prove Theorem \ref{thm:equiv}.

($\Leftarrow$) Given the uniqueness of $\m{T}$ satisfying \eqref{eq:HTLMI} and applying Lemme \ref{lem:embedding}, we have that there exist unique parameters $\lbra{s_k, f_k, \tau_k}_{k=1}^K$ satisfying \eqref{eq:Yexpression} and \eqref{eq:Texpression}. Then, it follows from \eqref{eq:Yexpression}, \eqref{eq:Yconstr} and the uniqueness of $\m{Y}$ that there exist unique parameters $\lbra{s_k, f_k, \tau_k}_{k=1}^K$ satisfying \eqref{2}, i.e., the parameters can be uniquely identifying from \eqref{2} (otherwise, it either contradicts the uniqueness of $\m{Y}$ or the uniqueness of the factorization in \eqref{eq:Yexpression}). Moreover, it follows from  \eqref{eq:Tgpsd} and Lemma \ref{lem:selective} that $\abs{\tau_k}\leq \delta$ for all $k$.

($\Rightarrow$) Since the parameters $\lbra{s_k, f_k, \tau_k}_{k=1}^K$ can be uniquely identified from $\m{y}$, applying Theorem \ref{thm:N2K}, we have $K\leq \frac{N}{2}<\frac{M+1}{2}$. Given the unique parameters $\lbra{s_k, f_k, \tau_k}_{k=1}^K$, define $\m{Y}$ and $\m{T}$ as in \eqref{2} and \eqref{eq:Texpression}. It is easy to verify \eqref{eq:Yconstr}. Moreover, \eqref{eq:HTLMI} and \eqref{eq:Tgpsd} follow from Lemma \ref{lem:embedding} and Lemma \ref{lem:selective}, respectively. 

Next, we show the uniqueness of $\m{Y}$ and $\m{T}$. Suppose that the $\m{Y}$ satisfying \eqref{eq:HTLMI}--\eqref{eq:Yconstr} is non-unique. Note that we have shown that $K<\frac{M+1}{2}$. It follows from Lemma \ref{lem:embedding} that there must exist another set of parameters $\lbra{s_k, f_k, \tau_k}_{k=1}^K$ satisfying \eqref{eq:Yexpression}, implying that there exist two distinct pameraters $\lbra{s_k, f_k, \tau_k}_{k=1}^K$ satisyfing \eqref{2}, contradicting the unique identifiability of the parameters from $\m{y}$. Therefore, $\m{Y}$ must be unique and admit a unique factorization as in \eqref{eq:Yexpression}. It then follows from Lemma \ref{lem:embedding} that $\m{T}$ is also unique, completing the proof.

Here are some remarks on Theorem \ref{thm:equiv}. By Theorem \ref{thm:equiv}, we have actually reformulated chirp parameter identification as a problem of 2-D harmonic retrieval \cite{yang2016vandermonde}, in which the 2-D frequencies are given by the duo $\sbra{f_k,\tau_k}$, given partial samples and interval prior knowledge on one dimension of the 2-D frequencies. In particular, $\m{Y}$ in Theorem \ref{thm:equiv}, given as in \eqref{eq:Yexpression}, corresponds to the 2-D signal that is parameterized by the same set of parameters $\lbra{s_k,f_k,\tau_k}_{k=1}^K$ as $\m{y}$, and $\m{y}$ consists of partial samples of $\m{Y}$ on a parabola according to \eqref{eq:Yconstr}. The constraint \eqref{eq:HTLMI} imposes the spectral-sparse structure of $\m{Y}$ according to Lemma \ref{lem:embedding}. The constraint \eqref{eq:Tgpsd}, together with \eqref{eq:HTLMI}, imposes the interval prior on $\lbra{\tau_k}$ by Lemma \ref{lem:selective}. During the preparation of this work, we found that a similar perspective of 2-D harmonic retrieval was introduced in the recent preprint \cite{yang2025gridless} independently. While the paper \cite{yang2025gridless} is focused on convex optimization methods for chirp parameter estimation, the present work studies the fundamental chirp parameter identifiability problem and formulates it as an equivalent nonconvex matrix optimization problem. We will study how to solve the optimization problem and thus resolve the identifiability problem numerically in the ensuing section.


\section{Algorithm} \label{sec:algorithm}
To identify the parameters $\lbra{s_k, f_k, \tau_k}_{k=1}^K$ from $\m{y}$ in \eqref{2}, by Theorem \ref{thm:equiv} we can solve equivalently the following matrix feasibility problem:
\equa{&&\text{Find } \m{Y},\m{T}, \st \left\{ \begin{array}{l}\begin{bmatrix} \overline{\m{T}} & \overline{\cH\m{Y}} \\ \cH\m{Y} & \m{T}\end{bmatrix} \in \bS_+^K,  \\
		\m{T}_{\delta} \geq\m{0},  \\
		Y[n,n^2] = y[n], \quad n=0,\dots,N-1,  \end{array}\right. \label{fea} }  
where $\m{Y}$, $\m{T}$ and $\m{T}_{\delta}$ are as defined in Theorem \ref{thm:equiv}. Once a solution to \eqref{fea} is found numerically, we can compute the so-called Carath\'{e}odory-Fej\'{e}r decomposition of the 2-level Toeplitz matrix $\m{T}$ as in \eqref{eq:Texpression}, following \cite{yang2016vandermonde}, to determine $\lbra{f_k,\tau_k}$. After that, $\lbra{s_k}$ can be computed from \eqref{2} using a simple least squares method. In this paper, we will apply the convex iteration algorithm \cite{dattorro2010convex} to solve the above rank-constrained optimization problem.

\subsection{The convex iteration algorithm}
In this subsection, we introduce the convex iteration algorithm \cite{dattorro2010convex} that aims to solve the following rank and positive-semidefinite constrained feasibility problem:
\begin{equation}
	\text{Find } \bm{A}, \  \st  \bm{A} \in \mathcal{C},\  \bm{A} \in \bS_+^r, \label{17}
\end{equation}
where $\m{A}\in\bC^{n\times n}$, $\mathcal{C}$ is a convex set, and $r< n$ denotes the rank. The convex iteration algorithm is motivated by the following equivalent reformulation of \eqref{17}:
\begin{equation}
	\min_{\m{A},\m{W}} \left \langle \bm{A},\bm{W} \right \rangle, \  \st  \bm{A} \in \mathcal{C},\  \bm{A} \geq \bm{0},\ \bm{0} \leq \bm{W} \leq \bm{I}, \ {\rm trace} (\bm{W})= n-r,  \label{18}
\end{equation}
where $\m{I}$ denotes an identity matrix. Note that if \eqref{17} is feasible, then its feasible solution $\m{A}^*$, together with $\m{W}^* = \m{U}^*\m{U}^{*H}$ where $\m{U}^*\in\bC^{n\times (n-r)}$ is an isometry matrix whose column space is the eigen-space associated with the zero eigenvalues [of  multiplicity $(n-r)$] of $\m{A}^*$, forms an optimal solution to \eqref{18} with a zero optimal value. Based on this equivalence, the convex iteration algorithm turns to \eqref{18} and solves for $\m{A}$ and $\m{W}$ alternately, yielding the following iterate:
\lentwo{\equa{
		\m{A}^{(j)}
		&\leftarrow& \argmin_{\m{A}} \left \langle \bm{A},\bm{W}^{(j-1)} \right \rangle, \  \st  \bm{A} \in \mathcal{C},\  \bm{A} \geq \bm{0}, \notag\\ \m{W}^{(j)} &\leftarrow& \m{U}^{(j)}\m{U}^{(j)H}, \label{20} }
}where $\m{U}^{(j)}\in\bC^{n\times (n-r)}$ is composed of the $(n-r)$ eigenvectors associated with the smallest $(n-r)$ eigenvalues of $\m{A}^{(j)}$. Starting with $\m{W}^{(0)} = \m{I}$, the first iteration then becomes
\equ{\min_{\m{A}} \tr\sbra{\m{A}},\ \st  \bm{A} \in \mathcal{C},\  \bm{A} \geq \bm{0},\notag}
which is a convex relaxation of the original problem \eqref{17}. From the second iteration on, the low-rankness of $\m{A}$ is continuously enhanced by suppressing the power located in an $(n-r)$-dimensional subspace associated with the smallest $(n-r)$ eigenvalues. Evidently, the non-negative objective value of \eqref{18} decreases monotonically during the iterations and thus the convex iteration algorithm is guaranteed to converge. A feasible solution to \eqref{17} is found once the objective value of \eqref{18} drops to zero (within numerical precision, in practice).

We note that the convex iteration algorithm can also be interpreted using tail minimization \cite{lai2018spark} or truncated nuclear norm minimization \cite{zhang2012matrix}, while their apparent equivalences seem to be omitted elsewhere. 

\subsection{The proposed algorithm}
In this subsection, we derive an algorithm to solve \eqref{fea} by applying the convex iteration algorithmic framework. In particular, let $\m{A}$ in \eqref{17} denote the square, block matrix $\begin{bmatrix} \overline{\m{T}} & \overline{\cH\m{Y}} \\ \cH\m{Y} & \m{T}\end{bmatrix}$ of order $\frac{(M+1)[(M-1)^2+2]}{2}$. Moreover, the Hankel, Toeplitz and conjugate structures in the block matrix, as well as the second and third constraints in \eqref{fea}, are all convex that form the set $\cC$. Consequently, to solve \eqref{fea}, the convex iteration algorithm computes the $j$th iterate $\sbra{\bm{T}^{(j)},\bm{Y}^{(j)}}$ by solving
\begin{equation}
	\min_{\bm{T},\bm{Y}} \left \langle \begin{bmatrix} \overline{\m{T}} & \overline{\cH\m{Y}} \\ \cH\m{Y} & \m{T}\end{bmatrix},  \bm{W}^{(j-1)} \right \rangle,
	\st \left\{ \begin{array}{l}\begin{bmatrix} \overline{\m{T}} & \overline{\cH\m{Y}} \\ \cH\m{Y} & \m{T}\end{bmatrix} \geq 0,  \\
		\m{T}_{\delta} \geq\m{0},  \\
		Y[n,n^2] = y[n], \quad n=0,\dots,N-1,  \end{array}\right. \label{21}
\end{equation}
where $\m{W}^{(j)}$ is as defined in \eqref{20} with $\m{U}^{(j)}$ being a $\frac{(M+1)[(M-1)^2+2]}{2} \times \sbra{\frac{(M+1)[(M-1)^2+2]}{2}-K}$ matrix composed of the eigenvectors associated with the smallest $\sbra{\frac{(M+1)[(M-1)^2+2]}{2}-K}$ eigenvalues of the latest block matrix $\begin{bmatrix} \overline{\m{T}} & \overline{\cH\m{Y}} \\ \cH\m{Y} & \m{T}\end{bmatrix}$. Moreover, the first iteration, with $\m{W}^{(0)}=\m{I}$, becomes
\begin{equation}
	\min_{\bm{T},\bm{Y}} \tr\sbra{\m{T}},
	\st \text{the same constraints as in \eqref{21},\notag}
\end{equation}
which is a convex relaxation of \eqref{fea}.

\section{Numerical Results} \label{sec:numerical}

In {\em Experiment 1}, we test the tightness of the lower bound $N=2K$ of Theorem \ref{thm:N2K} by applying the proposed algorithm implemented with CVX and the SDPT3 solver on Matlab. Since it is non-tight as $K=1$ by Theorem \ref{thm:K1}, we consider the cases when $K=2,3,4$. We choose the parameters according to Table \ref{t1}. It is shown that the recovery errors of $\lbra{f_k,\tau_k}$ produced by the proposed algorithm are all on the order of $10^{-9}$ or lower, demonstrating that the lower bound $N=2K$ is attainable by the proposed algorithm and thus it is tight.

\begin{table*}[h!]
    \centering
    \begin{tabular}{ccc}
    \toprule
    & Ground truth of $\{s_k,f_k,\tau_k\}$ & Recovery errors of $\{f_k,\tau_k\}$  \\
    \cmidrule{1-3}
  \multirow{2}{*}{$K=2,\ N=4,\ \delta=0.05$} &  $\{e^{j\frac{\pi}{4}},0.25,0.02\}$ & $\{1.5\times10^{-9},4.9\times10^{-10}\}$ \\ 
  & $\{e^{j\frac{\pi}{6}},-0.3, 0.012\}$   & $\{1.4\times10^{-9}, 1.0\times10^{-9}\}$ \\
  \cmidrule{1-3}
  \multirow{3}{*}{$K=3,\ N=6,\ \delta=0.01$} &  $\{e^{j\frac{\pi}{4}},0,0.001\}$ & $\{3.3\times10^{-11},4.0\times10^{-9}\}$ \\ 
  & $\{e^{j\frac{\pi}{6}},0.33, 0.009\}$   & $\{4.4\times10^{-11},1.2\times10^{-10}\}$ \\
  & $\{e^{j\frac{\pi}{10}},-0.34, 0.005\}$   & $\{4.1\times10^{-11},1.0\times10^{-9}\}$ \\
  \cmidrule{1-3}
  \multirow{4}{*}{$K=4,\ N=8,\ \delta=0.01$} &  $\{e^{j\frac{\pi}{4}},0,0.001\}$ & $\{5.4\times10^{-13}, 3.8\times10^{-10}\}$ \\ 
  & $\{e^{j\frac{\pi}{6}},0.24,0.004\}$   & $\{7.7\times10^{-13}, 4.5\times10^{-9}\}$ \\
  & $\{e^{j\frac{\pi}{10}},0.49, 0.006\}$   & $\{1.2\times10^{-12}, 3.6\times10^{-9}\}$ \\
  & $\{e^{j\frac{2\pi}{5}},-0.25, 0.009\}$   & $\{1.2\times10^{-12}, 1.5\times10^{-10}\}$ \\
   \bottomrule
    \end{tabular}
    \caption{Parameter setup and absolute recovery errors by using the proposed algorithm.}
    \label{t1}
\end{table*}

\begin{figure*}[htbp]
	\centering
	\includegraphics[width=0.4\textwidth]{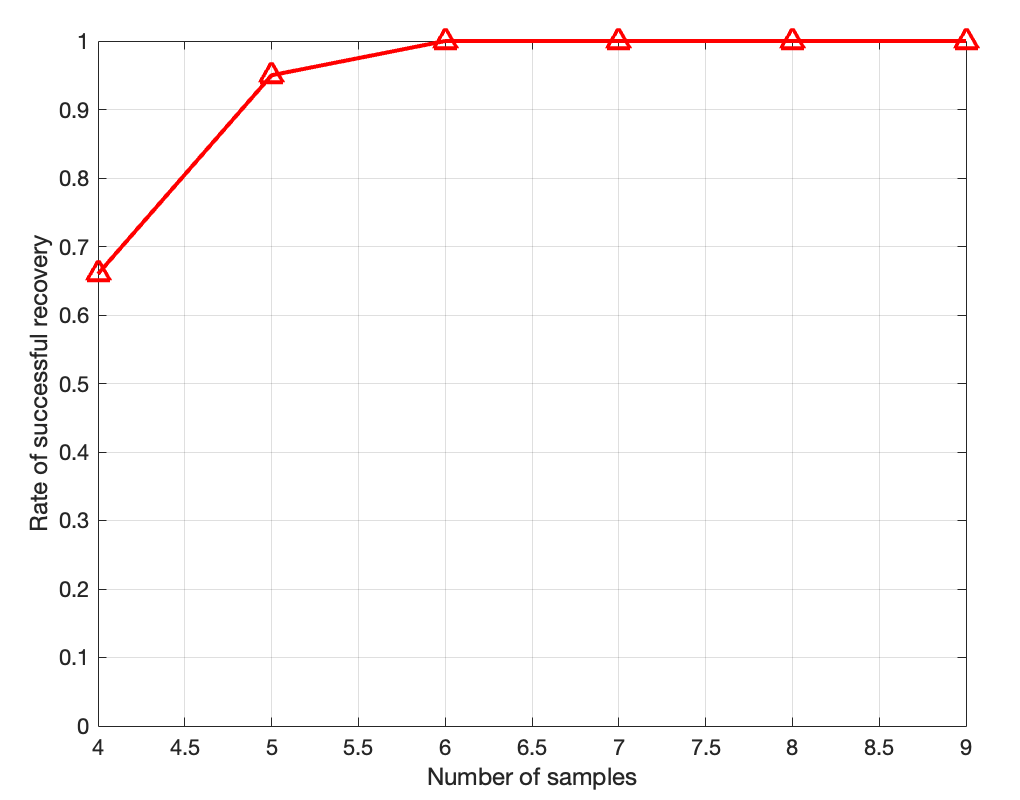}
	\caption{Rate of successful recovery of two chirps versus $N$.}
	\label{fig:rateofsucc}
\end{figure*}

In {\em Experiment 2}, we test the algorithm's performance with randomized parameters. In particular, we consider $K=2$ and randomly generate the parameters $\lbra{\abs{s_k},f_k,\tau_k}$ in small intervals which are centered at those used in {\em Experiment 1} and have length of $0.02$, $0.002$ and $0.2$, respectively. The phases of $\lbra{s_k}$ are uniformly distributed on $[0,2\pi]$. We consider the values of $N$ ranging from 4 to 9. Successful parameter identification is claimed if the absolute recovery errors of $\lbra{f_k,\tau_k}$ are all below $10^{-6}$. We repeat 100 times for each $N$ and compute the rate of successful recovery. Our numerical results are presented in Fig.~\ref{fig:rateofsucc}. It is seen that the rate reaches $66\%$ even at the minimal allowed value of $N=4$, verifying the good performance of the proposed algorithm in bound attainment. Moreover, the rate increases constantly as $N$ gets larger, which is consistent with Theorem 
\ref{thm:largeN}, and no failures have been observed as $N\geq 6$. 

\begin{figure*}[htbp]
	\centering
	\includegraphics[width=0.4\textwidth]{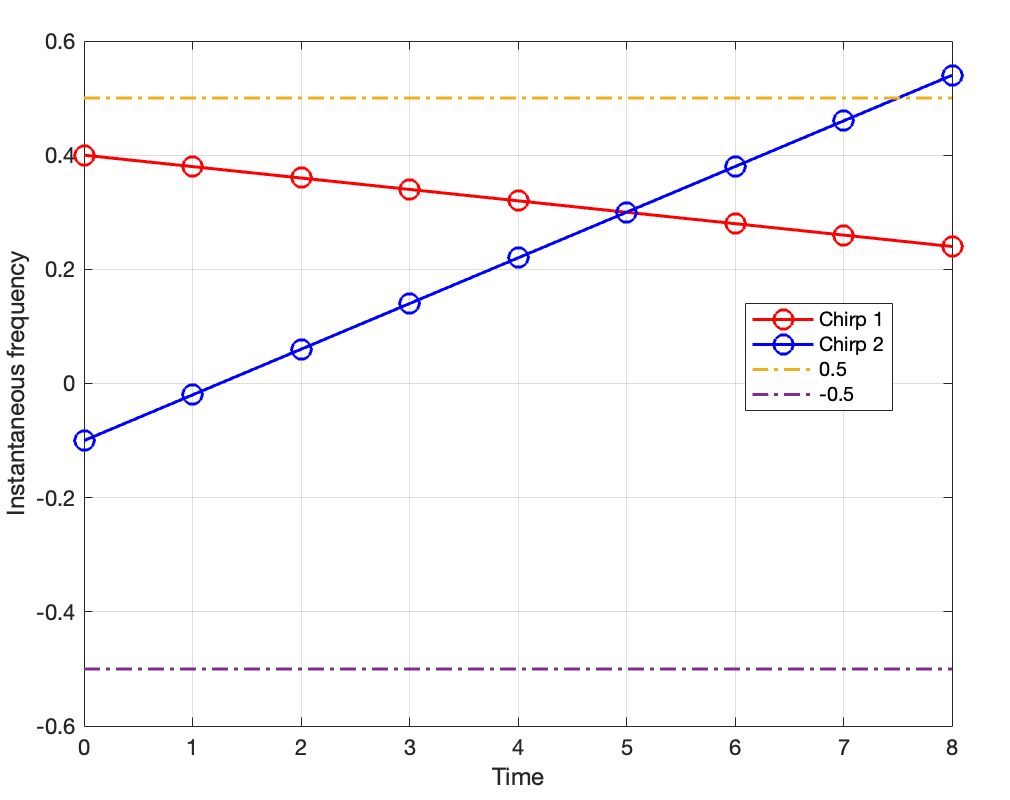}
	\caption{Plot of $K=2$ chirps in the time--frequency domain.}
	\label{fig:chirps}
\end{figure*}

%
%
In {\em Experiment 3}, we consider $K=2$ chirps with $\{s_1,f_1,\tau_1\}=\{e^{j\frac{\pi}{4}},-0.1,0.04\}$, $\{s_2,f_2,\tau_2\}=\{e^{j\frac{\pi}{6}},0.4,-0.01\}$, which are plotted in Fig.~\ref{fig:chirps} in the time--frequency plane. It is seen that the two chirps intersect at $n=5$. Moreover, the instantaneous frequency of one chirp exceeds $0.5$ at $n=8$, resulting in sub-Nyquist sampling. We assume $\delta=0.05$ and use $N=4,\dots,9$ samples, respectively, to identify the two chirps using the proposed algorithm. Again, the recovery errors of $\lbra{f_k,\tau_k}$ are all on the order of $10^{-9}$ or lower. We only present the errors with $N=4,9$ in Table \ref{t2} for conciseness. This verifies that the parameters can be successfully identified for crossing chirps or with sub-Nyquist sampling using the proposed algorithm.

\begin{table*}[h!]
	\centering
	\begin{tabular}{cc}
		\toprule
& Recovery errors of $\{f_k,\tau_k\}$  \\
		\midrule
		\multirow{2}{*}{$N=4$} 
& $\{1.3\times10^{-9},\ 4.7\times10^{-10}\}$ \\
		& $\{1.3\times10^{-9},\ 4.0\times10^{-10}\}$ \\
		\cmidrule{1-2} 
		\multirow{2}{*}{$N=9$} 
 & $\{6.8\times10^{-11},\ 1.3\times10^{-12}\}$ \\
 & $\{2.1\times10^{-11},\ 1.7\times10^{-12}\}$ \\
\bottomrule
\end{tabular}
\caption{Absolute recovery errors of $\lbra{f_k,\tau_k}$ with $K=2$ chirps and different choices of $N$.}
	\label{t2}
\end{table*}


\begin{figure*}[htbp]
\centering
	\subfigure[]{\includegraphics[width=0.3\textwidth]{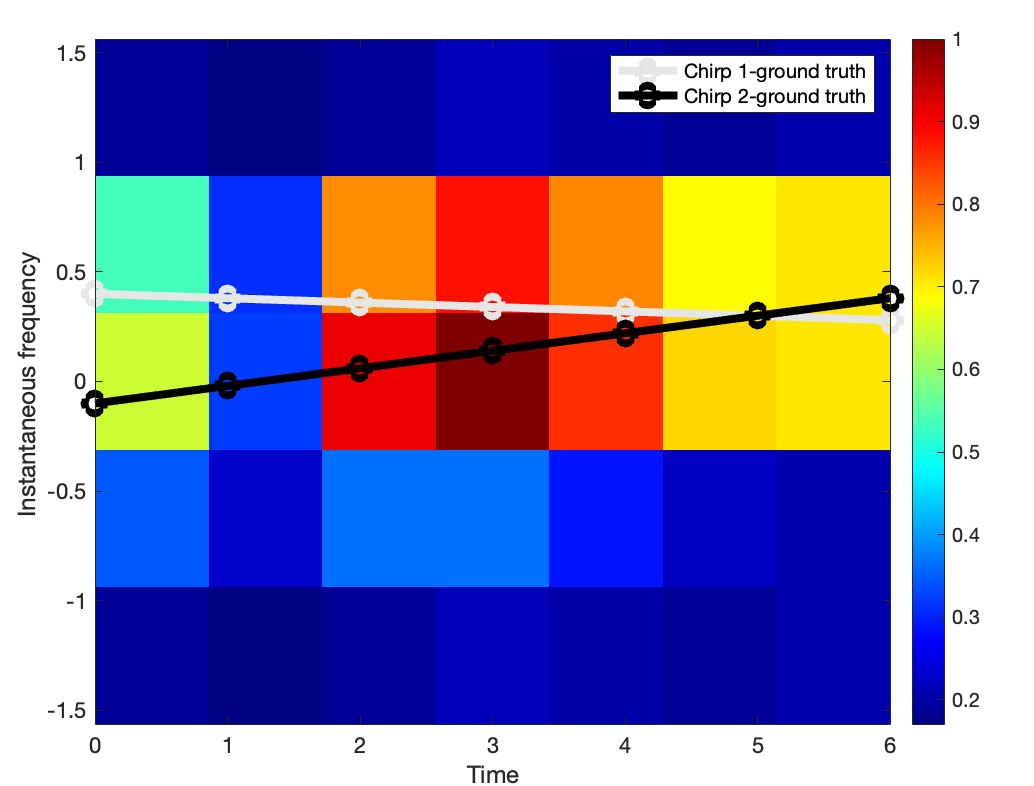}} 
	\hfill
	\subfigure[]{\includegraphics[width=0.3\textwidth]{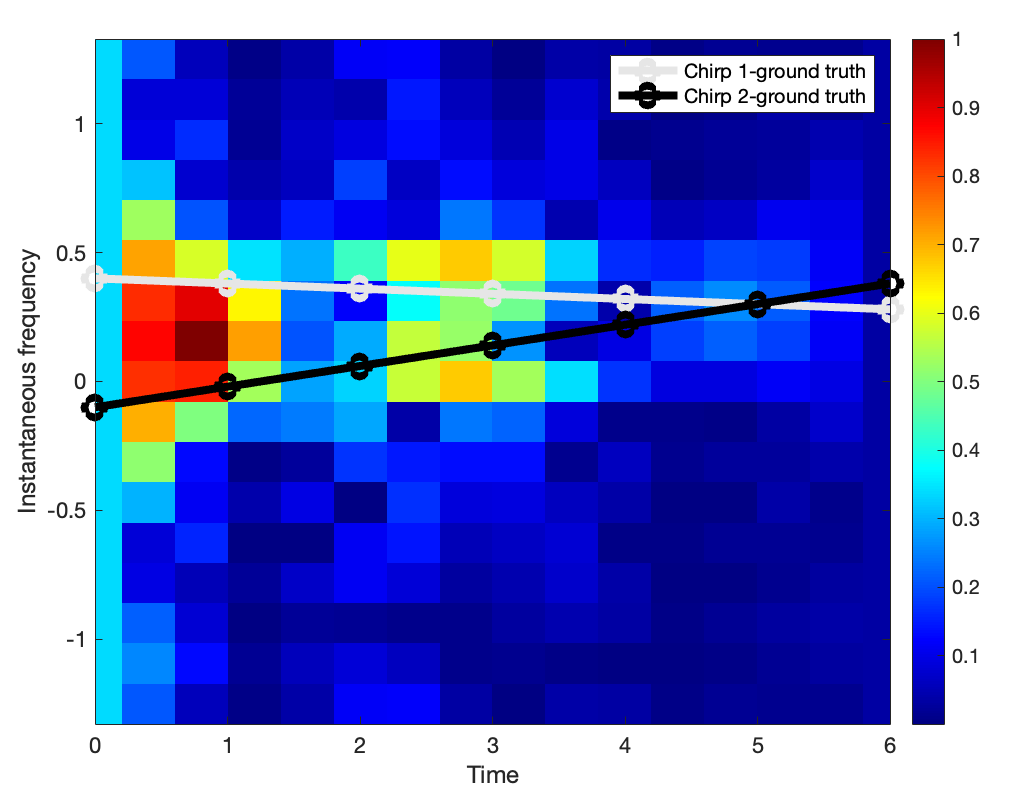}}
	\hfill
	\subfigure[]{\includegraphics[width=0.3\textwidth]{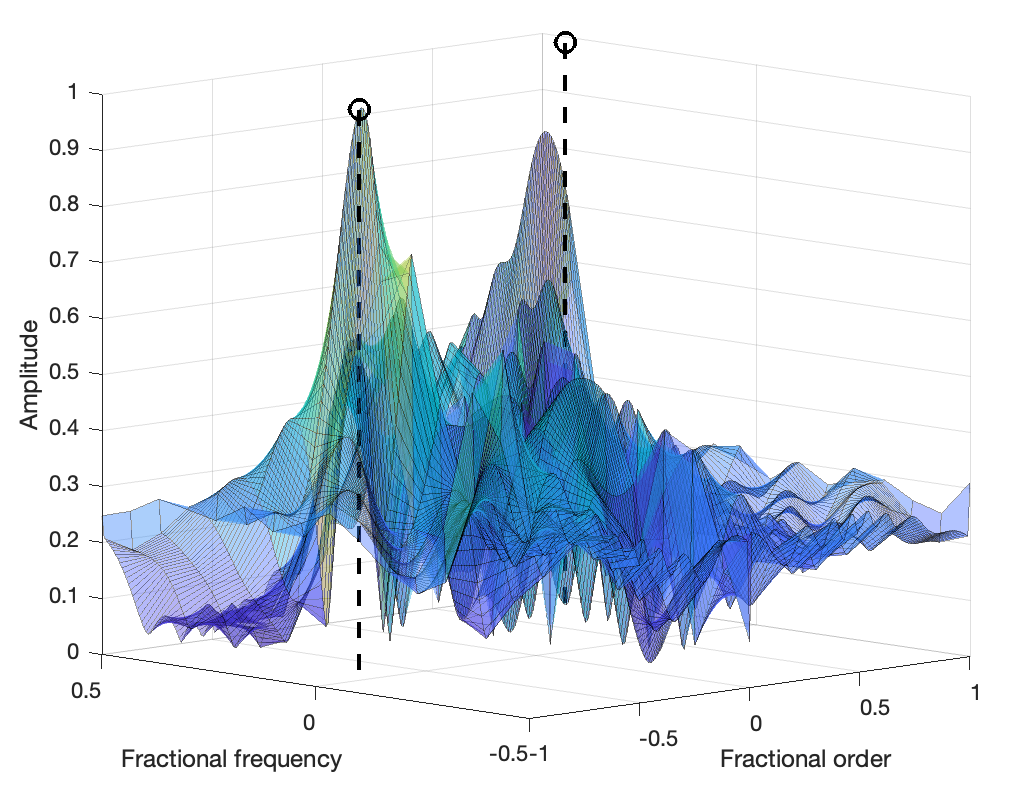}}\\
	\hfill
	\subfigure[]{\includegraphics[width=0.3\textwidth]{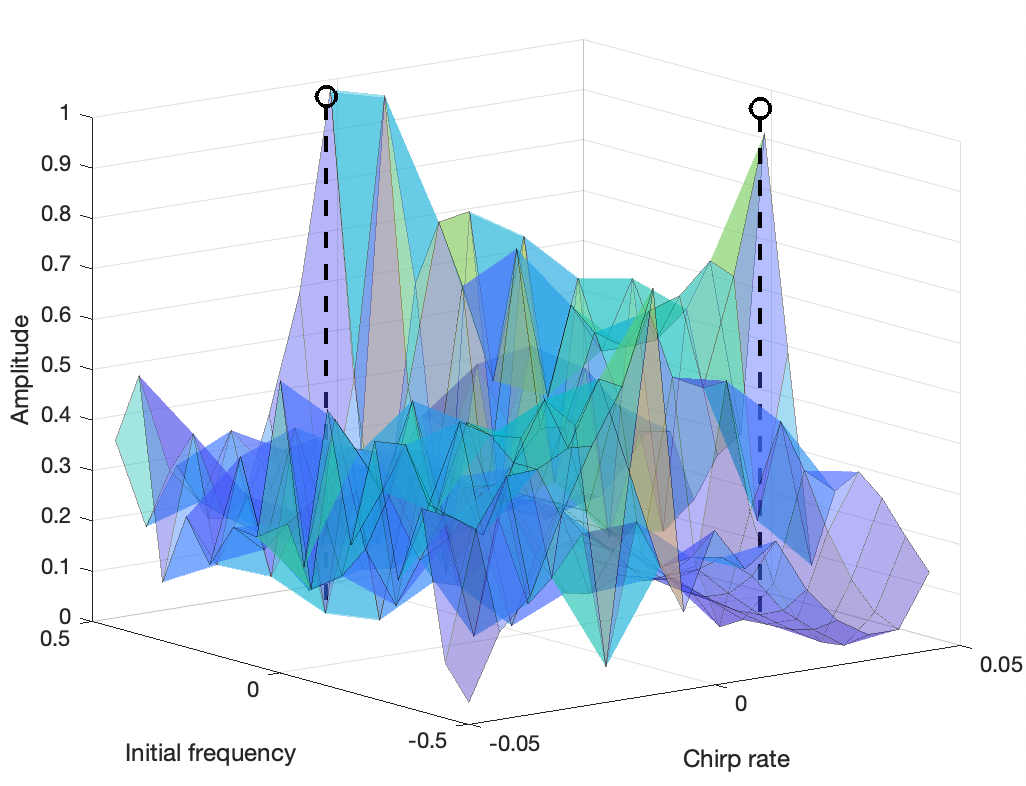}}
	\hfill
	\subfigure[]{\includegraphics[width=0.3\textwidth]{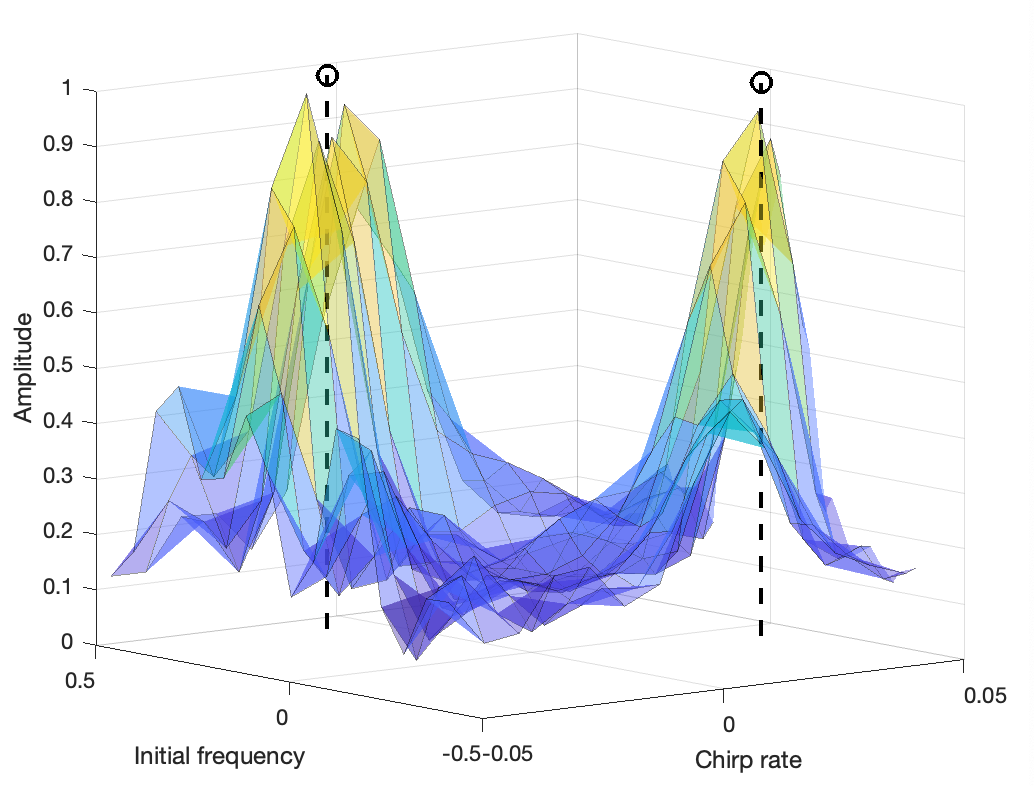}}
	\hfill
	\subfigure[]{\includegraphics[width=0.3\textwidth]{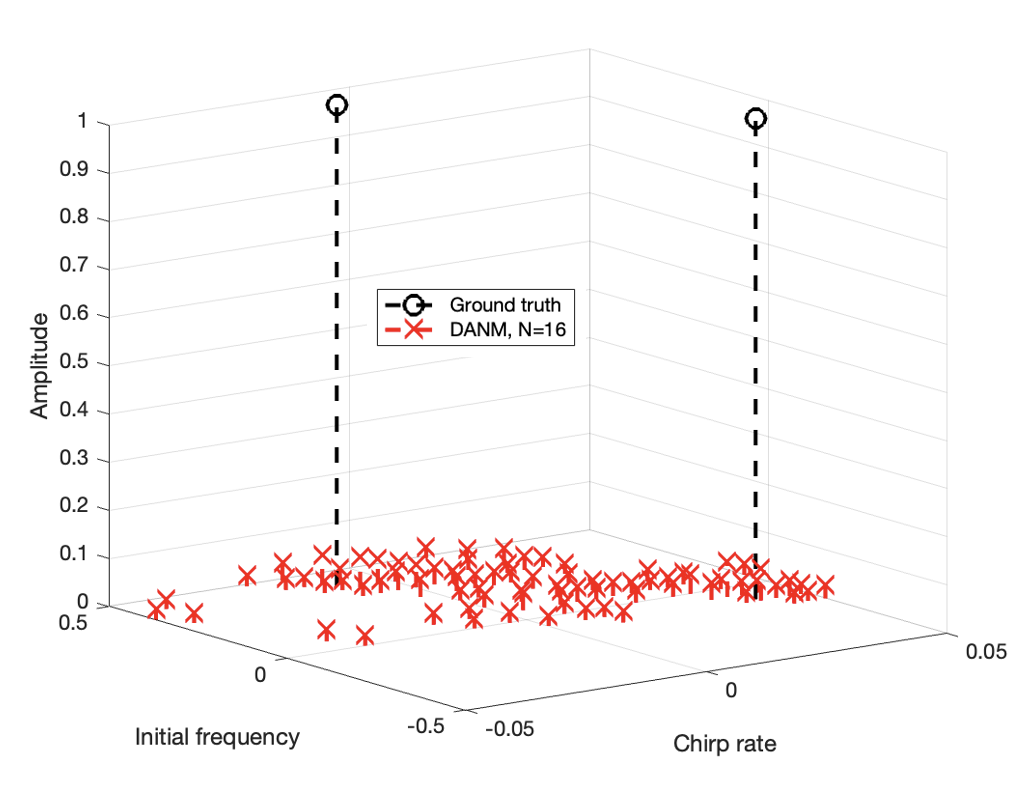}}
\caption{Results of identifying two chirps using (a) STFT, (b) WVD, (c) FRFT, (d) DCFT, (e) LVD and (f) DANM with 16 samples. It is shown in Table \ref{t2} that the two chirps can be exactly identified using the proposed algorithm with 4 samples only.}	\label{fig3}
\end{figure*}

In {\em Experiment 4}, we compare the proposed algorithm with existing methods including STFT \cite{tao2009short}, WVD \cite{debnath2002wigner}, FRFT \cite{ozaktas1996digital}, DCFT \cite{fan2000modified}, LVD \cite{lv2011lv} and decoupled atomic norm minimization (DANM) \cite{yang2025gridless}. In particular, we try to identify the two chirps used in {\em Experiment 3} by using these algorithms. While it is shown in {\em Experiment 3} that $N=4$ samples are sufficient to identify the $K=2$ chirps using the proposed algorithm, no meaningful results can be obtained using the above-mentioned methods in this case. To obtain more meaningful results with these methods, we increase the sample size from $4$ to $16$; to be specific, for STFT, WVD, FRFT, DCFT and LVD, the time duration and sampling frequency are twice and $2.5$ times those for the proposed algorithm respectively, and for DANM the time duration is increased to 5 times that for the proposed algorithm with a same sampling frequency. The obtained results are presented in Fig.~\ref{fig3}. It is shown that the chirp parameters still cannot be accurately identified by these methods, demonstrating the superior performance of the proposed algorithm.

\section{Conclusions} \label{sec:conclusion}
In this paper, the limit of chirp parameter identifiability was studied from a computational perspective. A necessary and sufficient condition is provided by deriving a rank-constrained matrix feasibility problem. An algorithm is proposed to solve the optimization problem, with which the lower bound $N=2K$ on the sample size $N$ with $K$ chirps is shown to be tight for parameter identifiability. The proposed algorithm is shown to significantly surpass the existing methods in terms of bound attainment.

The results of this paper can be extended in several ways in future studies. For example, it is of practical interest to extend the optimization problem and algorithm to the noisy case. While the chirp parameter estimation has been cast in this paper as the problem of 2-D harmonic retrieval with a specialized sampling pattern, it is also interesting to extend the results to general harmonic retrieval in arbitrary dimension. 



\end{document}